\documentclass[aps,groupedaddress]{revtex4}
\usepackage{epsfig}
\usepackage{graphicx}
\usepackage[T1]{fontenc}
\usepackage{ae}
\usepackage{color}
\usepackage[latin1]{inputenc}
\usepackage{amssymb,amsbsy,amsmath}
\usepackage{bbm}

\newtheorem{lemma}{Lemma}

\newtheorem{prop}{Proposition}

\newtheorem{prin}{Principle}

\newcommand{\kB}{k_{\rm B}}

\begin{document}



\title[Szilard engine]
{Coarse-grained entropy balance of the Szilard engine}

\author{Heinz-J\"urgen Schmidt and Thomas Br\"ocker
}
\affiliation{ Universit\"at Osnabr\"uck,
Fachbereich Mathematik, Informatik und Physik,
 D - 49069 Osnabr\"uck, Germany
}


\begin{abstract}
In order to reconcile the entropy reduction of a system through external interventions
that are linked to a measurement with the second law of thermodynamics, there are two main proposals:
(i) The entropy reduction is compensated by the entropy increase as a result of the measurement on the system (``Szilard principle").
(ii) The entropy reduction is compensated by the entropy increase as a result of the erasure of the measurement results (``Landauer/Bennett principle").
It seems that the LB principle is widely accepted in the scientific debate.
In contrast, in this paper we argue for a modified S principle and criticize the LB principle with regard to various points.
Our approach is based on the concept of ``conditional action", which is developed in detail.
To illustrate our theses, we consider the entropy balance of a variant of the well-known Szilard engine, understood as a classical mechanical system.
\end{abstract}

\maketitle

\section{Introduction}\label{sec:I}

Since its conception in 1867, James Clerk Maxwell's thought experiment has inspired
countless ideas and probably spawned over a thousand scientific papers \cite{footnote}.
In this famous scenario, an intelligent and skillful being
(later called a ``demon" by William Thomson)
controls a small door between two gas chambers.
When individual gas molecules approach the door,
the demon selectively opens and closes it.
This allows only the fast molecules to enter one chamber,
while the slower ones are directed into the other.
As a result, one chamber heats up while the other cools down,
which appears to reduce the entropy of the overall system and thus to violate the second law of thermodynamics.

Numerous works have attempted to rescue the second law by either debunking
Maxwell's thought experiment as impossible or discovering hidden entropy-increasing effects.
From the second group, the influential works of Szilard \cite{S29},
Landauer  \cite{L61}, and Bennett \cite{B82} should be mentioned.
Szilard introduced a refined version of Maxwell's thought experiment,
known as ``Szilard's engine", in which a single gas molecule is enclosed in a cylindrical box divided by a piston.
Depending on the position of the molecule - either in the left or in the right chamber -
an isothermal expansion to the right or to the left is carried out.
Apparently, heat is completely converted into work and thus the Szilard process
would constitute a {\it perpetuum mobile} of the 2nd kind (PM2).
Szilard argued that the entropy decrease of the heat bath is
compensated by the entropy cost of gaining information about the position of the molecule,
a concept known as ``Szilard's principle" (hereafter referred to as ``S principle'').

Building on Landauer's thermodynamic analysis of computation \cite{L61},
Bennett shifted the focus from the entropy cost of information
acquisition to the cost of information erasure \cite{B82}.
He pointed out that measurements can also be carried out reversibly,
i.~e.~without an increase in entropy.
Moreover, he claimed that for the cyclic operation of Szilard's engine
the memory that stores the respective measurement result of the molecule's position
must be reset to a default state each time.
This act of erasure inevitably generates entropy in the environment of the system,
which is at least as large as the entropy reduction achieved by the engine. This view,
known as ``Landauer/Bennett principle", has been widely accepted in the physics community.
However, it has also been criticized by a minority of scientists (\cite{EN98}, \cite{EN99}, \cite{N19}, \cite{N25} and others).
Recently R.~E.~Kastner has argued \cite{K25}, that the fundamental principle that saves the
second law is not the Landauer/Bennett principle but the quantum mechanical uncertainty principle,
which plays a role, for example, in the localization of the molecule in the Szilard engine.

For the purposes of this paper, it is useful to distinguish between the narrow interpretation of
``Landauer's principle" (L principle) which states that erasing memory content produces entropy
 - and the broader ``Landauer/Bennett principle" (LB principle),
which posits that this effect resolves the apparent paradox of Maxwell's demon.

The numerous works mentioned above on the paradox of Maxwell's demon
cover a wide range of different assumptions and premises.
Therefore, it is useful that we make explicit the decisions that underlie
our treatment of the paradox with a focus on Szilard's thought experiment:
\begin{itemize}
  \item
The demon was originally introduced as an intelligent being with more
refined observation and manipulation skills than humans.
Technical progress since Maxwell's time has made it possible to
actually carry out ``demon-like" experiments, see for example \cite{Ketal15}.
We can therefore replace the demon by a human experimenter or, in the next step,
by an apparatus with built-in measuring and control capabilities.
It is not necessary to develop a theory of intelligent beings that can make conscious
decisions in order to understand Maxwell's demon; one can ``naturalize" it.
\item
Many discussions of Maxwell's demon work with information-theoretic terms, 
see, e.~g., \cite{MNV09}.
It has even been proposed to extend the second law to include terms relating to ``information", see \cite{SU10}, \cite{SS15}.
Concerning this question it has been objected \cite{EN99}, \cite{N25} that information-theoretic approaches
are inadequate because a consistent theory for them is not yet available.
We can leave this question open, because in our solution approach we only want
to use physical terms in the narrower sense and not information-theoretical terms.
We also include ``Shannon entropy" and ``mutual information" among the physical
terms in the narrower sense, although this point could be debated.
It could not be ruled out that our approach can also be interpreted in terms of information theory;
but we do not need an extension of known physics to analyze Maxwell's demon.
\item
This gives rise to the task of developing a physical model for demonic interventions
or, as we will call it more neutrally, ``conditional actions".
But which theories should be used for this? Since we are dealing with
measurements on individual molecules, quantum mechanics offers
itself as a theoretical framework. But a purely classical description could also be interesting.
On the one hand, two types of problems can then be separated:
Problems of modeling Maxwell's demon from unsolved problems of quantum mechanics,
such as the formulation of a second law or the quantum mechanical measurement problem.
On the other hand, there is an extensive class of classical
``sorting processes" in the generalized sense,
which are not based on microscopic measurements and have entropy-reducing effects.
We will therefore focus on classical conditional actions and
only mention some other work \cite{Z84}, \cite{BS95}, \cite{Ketal11}, \cite{A13}, \cite{S20}, \cite{K25}
that deals with quantum mechanical versions of Szilard's machine.
\item
As already mentioned, the paradoxical character of Szilard's process
can be seen in the fact that it represents a possible PM2,
i.~e.~a cyclic process in which work is gained by 
extracting heat from a heat bath
while leaving everything else unchanged.
Such a process would reduce the total entropy of a closed system and
thus violate the 2nd law.
When it comes to the question of why the Szilard process does not
represent a PM2 on closer inspection, the representatives of the
various principles are largely in agreement:
it is due to the altered state of the memory for the measurement result.
The differences between the principles mentioned, however,
arise from the entropy balance of the individual phases of the Szilard process,
which we will therefore examine in more detail.
\item
Of course, the question then arises as to which entropy concept to use.
Here we have opted for a coarse-grained entropy concept by dividing
the entire phase space of the system into a finite number of cells
and calculating the Shannon entropy in terms of the probabilities
with which these cells are occupied.
\end{itemize}

A similar approach to the one outlined above was pursued in the work of P.~N.~Fahn \cite{F96}.
Here, too, an entropy balance was carried out for the individual phases of a classical Szilard process,
which methodically agrees with our approach for the special case of uniform
distributions in the spatial part of the molecule's phase space.
However, Fahn considers a different protocol of the Szilard process,
which is based on that of Bennett \cite{B87}, and which is not interesting
from our point of view because there are no entropy reductions of the gas at all
to be explained.

The conceptual reduction to classical systems with a finite number of states
makes it possible to explain the entropy reduction by ``pure" conditional actions
in a rather elementary way:
A pure conditional action generates a mapping $f:\Sigma \to \Sigma$ of the set of states of the system
$\Sigma$
into itself; each such mapping decreases the Shannon entropy of the system
or leaves it constant in the special case where $f$ is a permutation.
The decrease of entropy through pure conditional action
thus goes hand in hand with a partial merging of states.
In the extreme case, in which $f$ maps all states of $\Sigma$ to
one and the same state, the Shannon entropy is reduced to its minimum value $0$.
A special case of the latter is the erasure of a memory content.

On the other hand, every pure conditional action can be realized reversibly,
i.~e.~by a permutation of the states of an extended system
$\Sigma {\mathcal A}$. Although this is an almost trivial statement of set theory,
the more challenging task is to realize the conditional effect
through a physical model of an apparatus ${\mathcal A}$ and the
interaction between $\Sigma$ and ${\mathcal A}$.
For the Szilard engine, such a model is outlined in sections
\ref{sec:P} and \ref{sec:MSE}.

In order to apply this approach to the Szilard process,
we need to modify it so that the conditional action phase with
merging of states can be clearly identified.
However, it turns out that the concept of a ``pure" conditional action
is too narrow to describe this application to the Szilard engine
and must be generalized to more general conditional actions
mediated by stochastic matrices, see subsection \ref{sec:MGA}.

The paper is structured as follows. After this Introduction
we outline in section \ref{sec:P} the different phases of the
Szilard process, modified for our purposes, and provide a first rough
entropy balance for these phases. The different principles of
Szilard, Landauer and Landauer/Bennett are evaluated with respect to this entropy balance.
Then, in Section \ref{sec:GTCA}, we turn to the general theory of conditional
actions for classical finite systems. After some generalities in subsection \ref{sec:L}
the definition of conditional actions is given in the subsections \ref{sec:CA} and  \ref{sec:BA}
and illustrated for the case of Maxwell's demon, see subsection \ref{sec:CAMD}.
The subsections \ref{sec:MBA} and  \ref{sec:MGA} deal with (abstract) physical realizations of conditional actions
and the formulation of a modified S principle. The modification essentially consists
in taking into account the mutual information between the system $\Sigma$
and the apparatus ${\mathcal A}$,
by which the conditional action is realized.
Section \ref{sec:MSE} is devoted to a physical realization of the conditional
action phase of the Szilard process defined in Section \ref{sec:P}, see subsection \ref{sec:GD},
and to the detailed entropy balance of this phase, see subsection \ref{sec:EB}.
The corresponding calculations confirm the rough balance given in Section \ref{sec:P}.
In subsection \ref{sec:EBM} we sketch another, three-boxed modification of the Szilard engine
with non-vanishing mutual entropy thus illustrating our modified S principle.
We close with a summarizing discussion in Section \ref{sec:SUM}.

\section{Szilard engine and discussion of various priciples}\label{sec:P}

\begin{figure}[ht]
  \centering
    \includegraphics[width=1.0\linewidth]{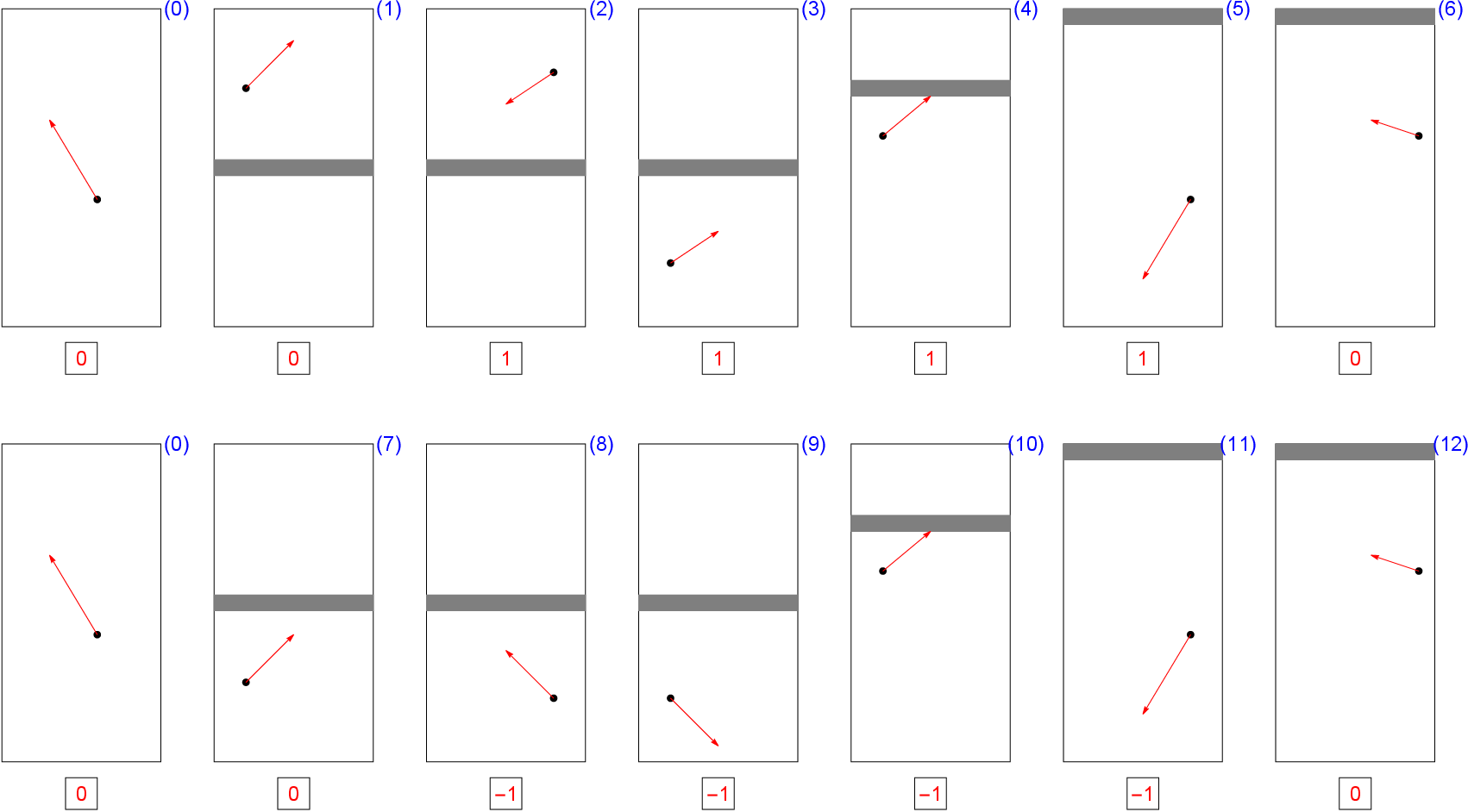}
  \caption{
  Sketch of the different phases of a possible Szilard engine cycle (modified from \cite{S29} and \cite{B87}).
  The whole container is put into a vertical gravitational field. The walls of the container serve as a heat bath
  for the one-molecule gas.  The memory is set to its default value $0$ (picture (0)).
  A partition is inserted which divides the container into two boxes of equal volume.\\
  Upper panel:
  The molecule is initially in the upper box (picture (1)).
  Then a position measurement of the molecule is made and its result $1$, which means ``upper box",  is stored in the memory (picture (2)).
  Depending on the result of this measurement a conditional action is performed:
  In the case where the particle has been found in the upper box,
  the container is rotated about $180^\circ$
  such that now the molecule is in the lower box (picture (3)). The partition is unlocked and serves as a piston that is moved upwards
  by the collisions with the molecule (picture (4)) until it reaches its final position (picture (5)). Then the
  memory is erased and set to the default value $0$ (picture (6)).\\
  Lower panel:  The molecule is initially in the lower box (picture (7)).
  Then a position measurement of the molecule is made and its result $-1$, which means ``lower box",  is stored in the memory (picture (8)).
  Depending on the result of this measurement a conditional action is performed:
  In the case where the particle has been found in the lower box, nothing is done (picture (9)).
  The partition is unlocked and serves as a piston that is moved upwards
  by the collisions with the molecule (picture (10)) until it reaches its final position (picture (11)).
  Then the   memory is erased and set to the default value $0$ (picture (12)).
   }
  \label{FIGP2}
\end{figure}

If you insert a partition wall into a container with $N$ gas molecules
which divides the total volume $V$ into equal parts, you will not get
exactly $N/2$ molecules in each part, but small deviations $N_1,N_2$ from $N/2$, so that $N_1+N_2=N$.
This would lead to a small isothermal expansion if the partition wall is made movable
and thus becomes a piston. Can work be obtained from these fluctuations in the number of particles in a partial volume?

In his work \cite{S29}, Szilard simplified this situation to the extreme case
in which there is only one molecule in the container, i.e. $N=1$.
Szilard did not discuss the details of the mechanism in \cite{S29},
however, he made it clear that a position measurement of the molecule
and what we will call a ``conditional action" is involved.

To make our approach available, we will modify the Szilard engine
in the following way, see Figure \ref{FIGP2}.
First,
we position the gas container in a gravitational field $(0,0,-g)$ parallel to the $z$-axis
in such a way that the work output during isothermal expansion
can be obtained by moving the piston upwards. The partition wall should be inserted at $z=0$,
which therefore also represents the initial position of the piston.
However, upward expansion only occurs if the molecule is in the lower box.
Therefore, we make the gas container rotatable around the y-axis and rotate it by $180^\circ$
if the molecule is initially in the upper box. This defines our conditional action.
In both cases, the molecule is in the lower partial volume after the conditional action
and can push the piston upwards. During the expansion phase,
the rotation of the gas container naturally remains blocked.
Once the expansion to the original volume of the gas is complete,
the piston can be lowered to the position in the center of the container,
releasing the potential energy gained to an external energy storage device.
Then the piston is again inserted as a partition wall of the container,
and the entire process can be continued cyclically.

This (modified) Szilard process is therefore a candidate for a
{\it perpetuum mobile} of the  2nd kind (PM2).
This term usually refers to a thermodynamic cyclic process
in which heat is completely converted into work,
whereby the overall state of the system + environment does not change,
except for negligible decrease in temperature of the heat bath from which the
converted heat originates and the content of an external energy store that absorbs the work gained.
During a cycle of a PM2, the entropy of the system would therefore remain constant,
while the entropy of the heat bath would decrease.

Thus, a PM2 violates the second law. The impossibility of a PM2
is often regarded as an equivalent formulation of the 2nd law
that does not require an explicit entropy definition.

We can leave the question of equivalence open
here, because we are only interested in whether
the Szilard process violates the 2nd law or not.
In this context, arguments concerning PM2 and arguments concerning the entropy balance
are usually mixed, also in \cite{S29}. We will try to separate these two types of arguments.

When we examine the question of whether the Szilard process is a PM2,
the memory in which the result of the position measurement is stored plays a role.
We assume that this memory initially displays the number $0$ as the default value
and then, after the position measurement, either the number $1$ for ``upper box"
or $-1$ for ``lower box", see Figure \ref{FIGP2}.
Since the memory must be counted as part of the environment of the  ``gas" system,
its state is different at the beginning and end of the supposed cycle of a process
and therefore the definition of a PM2 is violated.

This is actually all that is needed to resolve the apparent paradox
represented by the Szilard engine, without invoking specific principles.
Additional arguments only come into play, when a proponent
of Szilard's PM2 proposal objects that the content of the memory could simply be erased at the end of the process.
Then one could rightly counter that the erasure of a memory content is by no means
thermodynamically innocent, but is only possible at the cost of an
entropy increase in the environment (L(andauer) principle).
(Although this is a balance argument, it is allowed here as an exception.)
We find that the L principle correctly supports the argument that the Szilard process is not a PM2.

This brings us to the part of the discussion that concerns the entropy balance.
Even though we have already ruled out the possibility that the
Szilard process as a PM2 lowers the overall entropy it is interesting
to see the detailed entropy balance of the individual phases of the Szilard process.
This is where the different positions of the S principle and the LB principle come into play.

For the entropy balance, it is important to define the system boundaries precisely.
For our approach, the following definition is useful:
The system (later referred to as $\Sigma {\mathcal A}$) comprises the one-molecule gas,
the memory and the entire setup required to operate the Szilard process,
including the heat bath necessary for isothermal expansion.
The rest of the world is referred to as ``environment";
it contains the energy reservoir to absorb the work done in the Szilard process.
In addition, entropy is released to the environment when the memory is erased,
i.~e.~possibly to a second heat bath, see section \ref{sec:GD}.

Then we claim the following results on the entropy changes $\Delta H$ of the various phases of the Szilard process,
see Figure \ref{FIGP2}.  $\Delta H$ is always understood as $\Delta H= H_{\scriptstyle{fin}}(\ldots)-H_{\scriptstyle{ini}}(\ldots)$
(with self-explaining notation)
and we refer to it as an {\em increase} if $\Delta H>0$ and a {\em decrease} if $\Delta H<0$.
Entropy is considered dimensionless by setting $\kB=1$, although the
Boltzmann constant will appear explicitly in some formulas in the usual way.
For the following we assume that the concept of mutual information is known
or refer to its later definition in subsection \ref{sec:L}.

\begin{enumerate}
  \item Insertion of the partition ((0) to (1) resp.~(0) to (7)): $\Delta H =0$.
  \item Measurement of the position of the molecule ((1) to (2) resp.~(7) to (8)): $\Delta H =0$. The entropy
  of the memory increases by $\log 2$ which is compensated by the subtraction of the
  mutual information between gas and memory of the same amount.
  \item Conditional action ((2) to (3) resp.~(8) to (9)): $\Delta H =0$. The entropy of the gas decreases by $\log 2$ which is compensated
  by the decrease of the mutual information of the same amount. After the conditional action the mutual information vanishes.
  \item Isothermal expansion ((3) to (5) resp.~(9) to (11)): $\Delta H =0$. The entropy of the heat bath  decreases by $\log 2$
  and the entropy of the gas increases by the same amount.
  \item Erasure of the memory content: ((5) to (6) resp.~(11) to (12)): $\Delta H =- \log 2$. The decrease of entropy
  is compensated by an increase of entropy of at least the same amount of the environment.
  \item Summarizing the single steps we obtain an entropy flow of $\log 2$ from the heat bath to the environment.
\end{enumerate}
The Szilard engine hence does not represent a PM2,
but rather a heat engine working between two temperatures (if the environment is modelled by a second heat bath
with a lower temperature, see subsection \ref{sec:GD}).

This detailed entropy balance enables us to evaluate the different points of view
with regard to the resolution of Maxwell's demon paradox.
At first sight, the constant entropy during the position measurement in phase (2)
seems to support the criticism of the S principle,
which identifies the alleged entropy increase by the measurement as
compensation for the entropy decrease of the gas.
However, if you take a closer look, you can see that there is no decrease in
entropy of the gas in phase (2) that needs to be explained.
This only takes place in phase (3), the conditional action.
It is also not clear whether the two phases (2) and (3) can always be separated,
as we have assumed in the diagram in Figure \ref{FIGP2}.
In the physical realization, which we will propose in section \ref{sec:MSE},
both phases are merged into one process.
So if we combine the two phases, measurement and conditional action,
into one phase, say phase (23), the entropy decrease of the gas is compensated
by the entropy increase of the memory, in accordance with the S principle.
However, we must make the following restriction:
The vanishing of the mutual information at the end of phase (23)
is a peculiarity of the Szilard engine; in general,
this mutual information will not vanish and the S principle
must be modified accordingly, see section \ref{sec:MBA}.

Furthermore, the above entropy balance casts a critical light on the
LB principle, according to which the entropy production by erasing the memory content
compensates for the initial entropy decrease of the gas.
This is questionable for two reasons.
Firstly, this explanation is superfluous because the (modified) S principle
already saves the 2nd law. The production of entropy in the environment by erasing the memory content
does not generate the missing entropy, but only makes it visible, so to speak.
On the other hand, the LB principle would also not be sufficient
to explain the apparent violation of the 2nd law between phase (3) and phase (5),
i~.e.~during the possibly long-lasting phase of isothermal expansion.
In this phase, the entropy of the gas increases again,
but at the expense of the entropy of the heat bath.
If the entropy reduction of the gas by the conditional action were not already compensated for,
the 2nd law would be violated during the isothermal expansion.

\section{General theory of conditional actions}\label{sec:GTCA}

\subsection{Finite classical systems}\label{sec:L}
Basic concepts are a finite set $\Sigma$ of (pure) states (or ``elementary events")
and probability distributions
$p_i,\; i=1,\ldots, N$ with $N=\left| \Sigma \right|$, i.~e.~functions
$p:\Sigma \to [0,1]$ with $\sum_{i\in \Sigma}p(i)=1$. We use both notations
$p(i)$ and $p_i$  synonymously. Such a probability distribution $p$ is called
a ``state". ``Pure states" are characterized by $p_i=\delta_{ij}$,
i.e. all $p_i$ equal $0$ except for $p_j=1$ for a special $j\in \Sigma$.
All other states are called  ``mixed".

There are six possible elementary events for a throw with a die
that all have the same probability $p_i=1/6,\, i=1,\ldots,6,$
if the die is not loaded.
As a further example we consider the decomposition of the phase space for a molecule
into a finite number of cells $i\in \Sigma$ and $p_i$ as the averaged thermal
probability for a rarefied gas in a box with temperature $T$.
This is reminiscent of a ``lattice gas model", see \cite{S01}.

The (Shannon) entropy $H(p)$ for a state $p$ is given by
\begin{equation}\label{defshannon}
H(p)= - \sum_i p_i\, \log p_i
\end{equation}
where for $p_i=0$ the limit value $\lim_{x\downarrow 0}x \log x =0$
is to be used. By ``$\log$" we denote the natural logarithm,
although in the original definition \cite{S48} the logarithm to the base $2$ was used.
Because $\log p_i \le 0$, $H(p)$ is always positive or $0$. It can be shown that
$H(p)=0$ is equivalent to ``$p$ is pure state". Recall that we always consider
dimensionless entropies by setting $\kB=1$.

In the application to the above-mentioned
``lattice gas model" the Shannon entropy can be viewed as a kind of coarse-grained
entropy. This appears to make more physical sense than the definition via an integral
with probability densities, since an arbitrarily fine decomposition of the phase space
into cells contradicts the quantum mechanical uncertainty principle.
Of course, there remains a certain arbitrariness in the definition of
coarse-grained entropy due to the choice of cell size,
but this is largely eliminated in the case of entropy {\em differences}.

For the dice example it follows that
\begin{equation}\label{Hdice}
 H(p) = - \sum_{i=1}^6 \frac{1}{6} \log \frac{1}{6} = \log 6 = 1.791759\ldots
 \;.
\end{equation}

The time evolution of finite classical system $\Sigma$ will be represented
by the transition $p\mapsto p'$ given by
\begin{equation}\label{timevo}
  p_j'= \sum_{i=1}^N T_{ij}\,p_j, \quad \mbox{ for } j=1,\ldots,N
  \;,
\end{equation}
where $T$ is a {\it doubly-stochastic} $N\times N$-matrix, i.~e.~a matrix
with non-negative entries and all row and column sums equal to $1$.
It follows from this that entropy does not decrease over time:
By the Birkhoff-von Neumann theorem \cite{B46}, \cite{vN53}
the matrix $T$ is a convex sum of permutation matrices,
and the entropy is invariant under permutations of $\Sigma$ and a concave function
of probability distributions \cite{CT11}.
Thus we seem to have provided a simple proof of the 2nd law;
but of course it remains to be shown that the time evolution of a classical
many-body system is approximately given by a double stochastic matrix via coarse graining.

Next we consider the composition $\Sigma\,{\mathcal A}$ of two classical systems $\Sigma$ and ${\mathcal A}$.
The set of pure states of $\Sigma\,{\mathcal A}$ is the Cartesian product
$\Sigma\times{\mathcal A}$, i.~e.~the set of pairs $(i,\mu)$ with $i\in \Sigma$ and $\mu\in {\mathcal A}$.
The states of the composite system are again probability distributions
$P:\Sigma\times{\mathcal A}\to[0,1]$. In the special case of product probabilities $P=p\otimes q$,
i.e. $P(i,\mu) =p(i)\,q(\mu)$ with $p:\Sigma \to [0,1]$ and $q:{\mathcal A}\to [0,1]$
we speak of the composition of {\em independent} subsystems.

If we calculate the probabilities for events in the two subsystems
ignoring the respective residual system, we obtain the
so-called ``marginal distributions" $p=M_1(P)$ and $q=M_2(P)$:
\begin{equation}\label{rand}
 p(i):= \sum_\mu P(i,\mu), \, i\in \Sigma, \quad q(\mu):= \sum_i P(i,\mu), \, \mu\in {\mathcal A}
 \;.
\end{equation}
In the case of statistically independent subsystems, they agree with the initial probabilities.

In general, three different entropies can be defined:
\begin{eqnarray}
\label{H12}
  H(\Sigma, {\mathcal A}) &=& - \sum_{i\mu} P(i,\mu) \log P(i,\mu),\\
  \label{H1}
  H(\Sigma) &=&-\sum_i p(i) \log p(i),\mbox{ with } p(i) \mbox{ according to } (\ref{rand}), \\
  \label{H2}
  H({\mathcal A}) &=&-\sum_\mu q(\mu) \log q(\mu),\mbox{ with }q(\mu) \mbox{ according to } (\ref{rand})
  \,:
\end{eqnarray}
Interestingly, the entropy is only additive in the case of statistically independent subsystems.
In the general case, however, the following applies
\begin{equation}\label{add}
   H(\Sigma, {\mathcal A}) = H(\Sigma)+ H({\mathcal A}) - \underbrace{ H(\Sigma : {\mathcal A}) }_{\ge 0}
\end{equation}
with the non-negative ``mutual information" $H(\Sigma : {\mathcal A})$ implicitly defined by this equation,
see, e.~g.~\cite{CT11}.
The entropy of a composite
system is therefore generally smaller than the sum of the entropies of the subsystems.
The difference is equal to the mutual information and takes into account (with a negative sign)
the additional order or correlation between the two subsystems.

This can be analogized to the relativistic concept of mass:
The total mass of two bodies is the sum of the individual masses
minus the binding energy (divided by $c^2$).
In both cases, an interaction between two subsystems leads to a deviation from additivity.

\subsection{Conditional actions}\label{sec:CA}

An intervention in a finite classical system $\Sigma$ transforms the probabilities
according to $p \mapsto p'$, i.~e.~maps states to states.
It is a natural condition that this transformation should respect convex linear combinations.
This implies that a conditional action can be described by a stochastic $N\times N$-matrix $T$ such that
\begin{equation}\label{condact}
  p_i'= \sum_{j=1}^N T_{ij}\,p_j, \quad \mbox{ for } i=1,\ldots,N
  \;.
\end{equation}
Note that a stochastic square matrix, by definition, has only non-negative entries and
constant column sums equal to $1$.

A special case is a {\em pure} conditional action, defined by the condition
that it maps pure states to pure states. This is equivalent to the condition
that each column of $T$ consists of standard unit vectors (i.~e.~with zero entries except one entry $1$).
Equivalently, pure conditional actions can be described by maps
\begin{equation}\label{abb}
  f:\Sigma \to \Sigma
\;.
\end{equation}
This transforms the probabilities as follows:
\begin{equation}\label{abbW}
 p_j'= \sum_{i\in\Sigma, f(i)=j} p_i,\quad j\in \Sigma
 \;.
\end{equation}

In the following subsections, we will present the theory of pure conditional actions
in some detail, as these are easier to understand.
In particular, a pure conditional action will always decrease the entropy of the system
or keep it constant. A general transformation of the kind (\ref{condact}) can increase or decrease
the entropy, and hence the non-increase of entropy has to be additionally
included in the definition of  non-pure conditional actions.
However, non-pure conditional actions play a crucial role in our
realization of the Szilard engine, which is why we also have to deal with the general case.

\subsection{Pure conditional actions}\label{sec:BA}

In this subsection we only deal with pure conditional actions given by a map
$f:\Sigma \to \Sigma$.
In general, $f$ will not be a permutation of $\Sigma$, but maps different elementary events
$i$ onto the same event $j$. If one considers that the description of the system
$\Sigma$ by finitely many (pure) states already represents a kind of coarse graining,
one could also view the pure conditional action as a  ``second coarse graining".

As an example, we again consider the dice experiment with $p_i=1/6$ for $i=1,\ldots,6$.
The conditional action consists of turning over every dice that shows one of the numbers $1,2$ or $3$,
However, if it already shows the numbers $4,5$ or $6$, nothing is done.
The following therefore applies: $f(1)=f(6)=6$, $f(2)=f(5)=5$ and $f(3)=f(4)=4$,
as well as $p_1'=p_2'=p_3'=0$ and $p_3'=p_4'=p_5'=1/3$.
How does the entropy transform under this intervention?
At the beginning, $H(p)=\log 6 =1.791759\ldots$, see (\ref{Hdice}),
but after the intervention $H(p')=\log 3 = 1.09861\ldots < H(p)$.
This conditional action has therefore lowered the entropy of the system $\Sigma$.

This also applies in general to interventions that fulfill (\ref{abbW}) together with (\ref{abb}):
\begin{prop}\label{P1}

Let $ f:\Sigma \to \Sigma$ be a pure conditional action and
(\ref{abbW}) be the corresponding transformation of probabilities. Then $H(p') \le H(p)$
and $H(p') = H(p)$ iff $f$ is a permutation if restricted to
$\Sigma_+^{(p)}:=\{ i\in \Sigma\left|\right.p(i)>0\}$.
\end{prop}
The {\bf  proof} can be found in \cite{S20}, section 5. For the convenience of the reader
we will reproduce it here with slight variations.

Let us first assume $\Sigma=\Sigma_+^{(p)}$.
If $f$ is a bijection then the Shannon entropy is unchanged,
$H(p') = H(p)$. If $f$ is not a bijection then it can be written as a finite composition of maps
which map exactly two states onto the same one and are injective else.
Thus it suffices to prove $H(p') < H(p)$ for maps $f$ satisfying, say, $f(1)=f(2)=j$ and
$f$ being injective on the set $\{3,4,\ldots, n\}$ (the definition domain of $f$ consists of $n$ states with
positive probability which may be less than $N=|\Sigma |$). Note that $p'(j)=p(1)+p(2)$
and $p'(f(i))=p(i)$ for $i=3,4,\ldots,n$. It follows that
\begin{eqnarray}
\label{condactmon1}
 H(p') &=& -p'(j)\log p'(j) -\sum_{i=3}^{n} p(i) \log p(i) \\
 \label{condactmon21}
   &=& -p(1) \log(p(1)+p(2)) - p(2) \log(p(1)+p(2))-\sum_{i=3}^{n} p(i) \log p(i) \\
   \label{condactmon3}
   &<& -p(1) \log p(1) -p(2) \log p(2)-\sum_{i=3}^{n} p(i) \log p(i)\\
   \label{condactmon4}
   &=& H(p)
   \;.
\end{eqnarray}
In line (\ref{condactmon3}) we have used
\begin{equation} \label{condactmon5}
 \log (p(1)+p(2)) > \log p(1), \, \log p(2)
\end{equation}
which holds due to $p(i)>0$ for all $i=1,\ldots,n$ and $\log$ being a strictly increasing function.

Returning to the general case we first note that $H(p') \le H(p)$ remains valid
if the limit process $p_i \to 0$ is carried out for $i\in\Sigma_0^{(p)}:=\{ i\in \Sigma\left|\right.p(i)=0\}$.
However, from $H(p')=H(p)$ we can only conclude that $f$ acts as a permutation on $\Sigma_+^{(p)}$
because the action of $f$ on $\Sigma_0^{(p)}$ does not change the value of $H(p')$.
\hfill$\Box$\\

\begin{figure}[ht]
  \centering
    \includegraphics[width=0.5\linewidth]{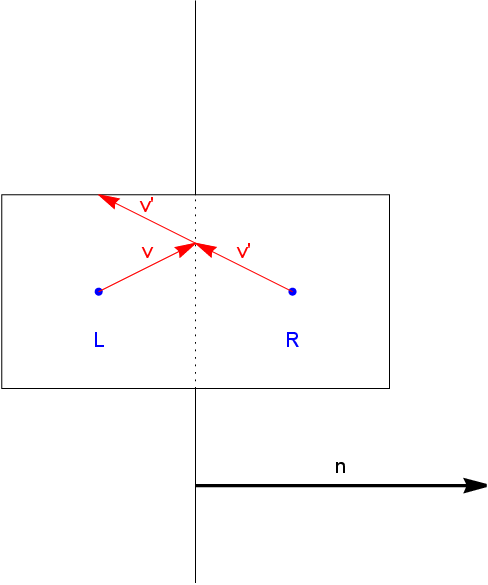}
  \caption{
  Sketch of the pure conditional action performed by Maxwell's demon.
  We show a part of the wall (with normal vector ${\mathbf n}$) separating two volumes filled with gas molecules
  and containing the small hole between the volumes.  The hole is covered by two cells denoted by $L$ and $R$.
  The further explanation can be found in the main text.
  }
  \label{FIGZ}
\end{figure}

\subsection{Conditional action by Maxwell's demon}\label{sec:CAMD}

We want to sketch how the interventions of Maxwell's demon can be
understood as pure conditional actions in the sense considered in the previous subsection \ref{sec:BA}.
It is assumed that the gas is so strongly rarefied that it can be described by a one-particle distribution.
The corresponding phase space is divided into $N= N_s\,N_p$ cells,
where $N_s$ cells account for the volume and $N_p$ cells for the momentum space.
Two spatial cells with the indices $R$ and $L$ cover the hole between the
partial volumes, see Figure \ref{FIGZ}.
We consider two cells in momentum space that are characterized by their mean vectorial velocities ${\mathbf v}$ and
${\mathbf v}'$, where ${\mathbf v}'$ corresponds to the velocity ${\mathbf v}$
reflected at the partition wall, that is
\begin{equation}\label{vrefl}
 {\mathbf v}' = {\mathbf v} - 2 {\mathbf v}\cdot {\mathbf n}\;{\mathbf n}
 \;,
\end{equation}
and ${\mathbf n}$ denotes the normal vector of the partition wall, see Figure \ref{FIGZ}.
In addition, we assume that the absolute velocity $\left|{\mathbf v}\right|=\left|{\mathbf v}'\right|$
belongs to a ``fast" molecule, which is allowed to pass from right to left, but not from left to right.

Then Maxwell's demon will perform a pure conditional action described by a
mapping $f:\Sigma \to \Sigma$, where $\Sigma$ denotes the finite set of cells defined above.
In particular the cell $(R,{\mathbf v}')$ will be mapped onto the cell $(L,{\mathbf v}')$
since the demon opens the door to let the fast molecule pass from right to left.
If, on the other hand, a fast molecule approaches the hole from the left
at a velocity of ${\mathbf v}$, the door remains closed and the molecule is reflected,
therefore $(L,{\mathbf v})$ is mapped onto the cell $(L,{\mathbf v}')$.
This is an example of two different cells being mapped onto the same cell by $f$.
The mapping $f$ is therefore not bijective, and, by virtue of proposition \ref{P1},
the entropy of $\Sigma$ decreases due to the pure conditional action of the demon.

\subsection{Models for pure conditional actions}\label{sec:MBA}

\begin{figure}[ht]
  \centering
    \includegraphics[width=0.5\linewidth]{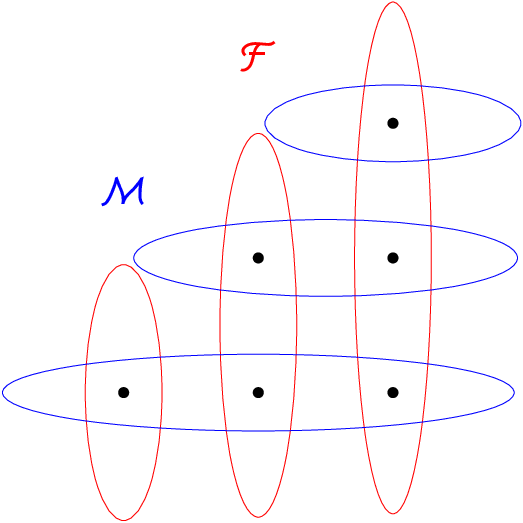}
  \caption{
  Sketch of two partitions ${\mathcal F}$ and ${\mathcal M}$ of the set $\Sigma$,
  the six elements of which are indicated by black dots.
  The ${\mathcal F}$-sets (red) and the ${\mathcal M}$-sets (blue) have at most one element in common.
  All elements of ${\mathcal F}$-sets are mapped onto the same element by the pure
  conditional action $f$, whereas all elements of ${\mathcal M}$-sets give the same result
  of the measurement linked with the conditional action.
  }
  \label{FIGPART}
\end{figure}

A pure conditional action given by a mapping $f:\Sigma \to \Sigma$ will
be modeled by the interaction with an apparatus ${\mathcal A}$, i.~e.~another physical system.
Of course, we have a model for the interaction of a naturalized Maxwellian demon in mind.

We consider the equivalence relation $i\sim j \Leftrightarrow f(i)=f(j)$ on $\Sigma$
and the corresponding partition ${\mathcal F}$ of $\Sigma$ into equivalence classes.
The equivalence class containing the element $i\in \Sigma$ will be denoted by $[i]_{\mathcal F}$.
Further let ${\mathcal M}$ be a second partition of $\Sigma$ satisfying
\begin{equation}\label{condFM}
 \left|G\cap M \right|\le 1 \quad \mbox{ for all } G\in {\mathcal F} \mbox{ and } M\in {\mathcal M}
 \;.
\end{equation}
Hence the ${\mathcal F}$-sets and the ${\mathcal M}$-sets have at most one element in common, see Fig.~\ref{FIGPART}.
The ${\mathcal M}$-set containing the element $j\in \Sigma$ will be denoted by $[j]_{\mathcal M}$.
The existence of such partitions  ${\mathcal M}$ is obvious (the axiom of choice being trivial for finite sets).
We think of the sets $[j]_{\mathcal M}$ of the partition ${\mathcal M}$ as the results of a measurement for the
state $j\in \Sigma$ and set ${\mathcal A}={\mathcal M}$,
thus reducing the apparatus to a memory.

Before the interaction begins, the memory is in the pure initial state $M_0\in {\mathcal M}$.
Therefore, the initial probabilities are of the product form
\begin{equation}\label{Pini}
 P_{\scriptstyle{ini}}(i,M) = p(i) \delta_{M M_0}\quad \mbox{ for all } i\in \Sigma \mbox{ and } M\in  {\mathcal M}
 \;,
\end{equation}
where the index ``{\it ini}" denotes the start of the conditional action and ``{\it fin}" its end.
We will describe the interaction by a permutation
\begin{equation}\label{WWPer}
 F:\Sigma\times {\mathcal M} \to \Sigma\times {\mathcal M}
 \;.
\end{equation}

This is the simplest ansatz, which ensures
that the total entropy remains constant during the interaction.
According to the above, doubly-stochastic transition matrices would also be conceivable,
but they would only further increase the entropy.

For initial states of the form $(i,M_0)$, the permutation $F$ should act like the
mapping $f$, therefore we require
\begin{equation}\label{postF}
 F\left(i,M_0\right)= (f(i),[i]_{\mathcal M})
\end{equation}
for all $i\in\Sigma$.
We will prove the following
\begin{lemma}\label{L1}
\begin{equation}\label{L1eq}
  F\left(i,M_0\right)=  F\left(j,M_0\right) \Rightarrow i=j \quad \mbox{ for all } i,j \in \Sigma
  \;.
\end{equation}
\end{lemma}
{\bf Proof}:
From $ F\left(i,M_0\right)=  F\left(j,M_0\right)$ and (\ref{postF}) it follows that
$G:=[i]_{\mathcal F}=[j]_{\mathcal F}$ and $M:=[i]_{\mathcal M}=[j]_{\mathcal M}$.
Since, according to (\ref{condFM}),  $G$ and $M$ have at most one common element
we conclude $i=j$.  \hfill$\Box$\\

This Lemma implies that we may extend $F$ to a
permutation on the whole set $\Sigma\times {\mathcal M}$ although this
extension is generally not unique. What exactly it looks like is irrelevant,
as long as we only consider the effect of $F$ on initial states $(i,M_0)$.
This effect can be characterized in such a way that $F$ changes the state $i$ of the system $\Sigma$
according to the conditional action $f$ and at the same time measures the state of $\Sigma$
and stores the measurement result in the memory ${\mathcal M}$.

For this model, we can concretize the entropy balance a little more.
Because the initial probability (\ref{Pini}) is of the product form,
the initial entropies are additive and the following applies
\begin{equation}\label{Hini}
 H_{\scriptstyle ini} (\Sigma, {\mathcal A})= H_{\scriptstyle ini} (\Sigma)+
 \underbrace{ H_{\scriptstyle ini} ({\mathcal A}}_{=0})= H_{\scriptstyle ini} (\Sigma)=
  -\sum_i p(i) \log p(i)
  \;,
\end{equation}
since at the beginning the apparatus was in a pure state $M_0$ with vanishing entropy.
After the end of the conditional action
\begin{equation}\label{Hfin}
  H_{\scriptstyle ini} (\Sigma)
  \stackrel{(\ref{Hini})}{=}
  H_{\scriptstyle ini} (\Sigma, {\mathcal A})
  =H_{\scriptstyle fin} (\Sigma, {\mathcal A})
   \stackrel{(\ref{add})}{=}
   H_{\scriptstyle fin} (\Sigma)+ H_{\scriptstyle fin} ({\mathcal A})- H_{\scriptstyle fin} (\Sigma : {\mathcal A})
  \;,
\end{equation}
where the constancy of the total entropy during the interaction was used for the second equality.
Hence the absolute value of the decrease of entropy $ \left|\Delta H(\Sigma) \right|$ of the system $\Sigma$ due to the conditional action
can be written as
\begin{equation}\label{DeltaH}
  \left|\Delta H(\Sigma) \right| =-\Delta H(\Sigma)= H_{\scriptstyle ini} (\Sigma) - H_{\scriptstyle fin} (\Sigma)
  \stackrel{(\ref{Hfin})}{=}
  H_{\scriptstyle fin} ({\mathcal A})- H_{\scriptstyle fin} (\Sigma : {\mathcal A})=
  \Delta H({\mathcal A})- H_{\scriptstyle fin} (\Sigma : {\mathcal A})
  \;.
\end{equation}
This means that the decrease of the entropy of $\Sigma$ is exactly compensated by the increase
of entropy of the apparatus ${\mathcal A}$ minus the mutual information $H_{\scriptstyle fin} (\Sigma : {\mathcal A})$.

\subsection{Models for general conditional actions and modified S principle}\label{sec:MGA}

Recall that a general, not necessarily pure, conditional action is described by a stochastic
matrix $T$ that transforms the probabilities according to
\begin{equation}\label{condact2}
  p_i'= \sum_{j=1}^N T_{ij}\,p_j, \quad \mbox{ for } i=1,\ldots,N
  \;.
\end{equation}

Let $p$ denote the mixed state with maximal entropy, i.~e., $p(i)=1/N$ for all $i\in\Sigma$.
Then the stochastic matrix $T$ is doubly stochastic iff $T p =p$. Hence, if $T$ is
not doubly stochastic, then $T p \neq p$ and hence $H(T p) < H(p)$ for this special
state $p$. But generally, it cannot be excluded that $T$ increases the entropy for certain states $p$.
As a physical example we note that the contact of a system with a heat bath with temperature $\theta$
(without assuming complete thermalization) can be described by a stochastic matrix $T$,
see \cite{SSG21}. Then we would obtain $H(Tp) < H(p)$ if the temperature of the system exceeds
$\theta$ and the system is cooled down by the contact, while $H(Tp) > H(p)$ in the other case
if the system's temperature is less then $\theta$.

Therefore, in the definition of conditional actions, 
we will additionally require that $H(p')\le H(p)$ holds, 
as this is not automatically fulfilled in the general case of stochastic matrices.

A {\em model} of the conditional action given by $T$ will be given by a triple $({\mathcal A}, { \rho},R)$.
Here ${\mathcal A}$ is a finite set of size $M$, ${\rho}$ is a probability distribution
${ \rho}: {\mathcal A}\to [0,1]$ and $R$ is a doubly-stochastic square matrix
with $N\times M$ rows. Moreover, the condition
\begin{equation}\label{condmod}
  T \, p = M_1\left( R\, p\otimes {\rho}\right)
\end{equation}
has to be satisfied for all probability distributions $ p: \Sigma \to [0,1]$,
where $M_1$ denotes the first marginal distribution and ${ p}\otimes {\rho}$
the initial product probability of the combined system $\Sigma {\mathcal A}$.

This definition is essentially identical with definition $1$ in \cite{S21b}.
A model of a conditional action hence consists of coupling $\Sigma$ to another system ${\mathcal A}$,
the ``apparatus", which is initially in the state ${ \rho}$,
such that the combined system $\Sigma{\mathcal A}$ undergoes a time evolution described
by $R$. The coordinate version of (\ref{condmod}) is
\begin{equation}\label{condmodcoor}
 \sum_n T_{m n} \, p_n = \sum_{ i j, n} R_{m i, n j}\, p_n\, \rho_j
 \;,
\end{equation}
for all $m=1,\ldots, N$ and hence the stochastic matrix $T$ can be
written in terms of $R$ and $\rho$ as
\begin{equation}\label{condmodT}
 T_{m n} = \sum_{ i j} R_{m i, n j}\,\rho_j
 \;,
\end{equation}
for all $m,n=1,\ldots, N$.

In \cite{S21b} it has been shown that every stochastic matrix $T$ admits a model in the sense
of the above definition. Especially one can set ${\mathcal A}=\Sigma$
and the initial state of the apparatus can be chosen as a pure state,
say, $\rho_{j} = \delta_{j 0}$ for all $j\in\Sigma$. We will not reproduce the proof given in \cite{S21b}
but only mention the special choice of the doubly-stochastic matrix $R$ made there:
\begin{equation}\label{defRdir}
  R_{mi,nj}:=\left\{
  \begin{array}{r@{\quad : \quad}l}
  T_{mi}\,\delta_{in} & j=0,\\
   \frac{1}{N(N-1)}\,(1-T_{mi}) & \mbox{else},
  \end{array}
  \right.
 \end{equation}
 for all $m,i,n,j\in{\Sigma}$.

Next we consider the entropy balance for a model of a general conditional action.
First we note that
\begin{equation}\label{balance1}
H_{\scriptsize{fin}}(\Sigma, {\mathcal A})= H(R\, p\otimes \rho) \ge H(p\otimes \rho) = H(p) + H(\rho)=
H_{\scriptsize{ini}}( \Sigma, {\mathcal A})
\;,
\end{equation}
where we have used the fact that application of a doubly stochastic matrix does not decrease the entropy.
Moreover, according to (\ref{add}),
\begin{eqnarray} \label{balance2}
 H_{\scriptsize{fin}}( \Sigma, {\mathcal A})&=& H\left(M_1(R\, p\otimes \rho) \right) +H\left(M_2(R\, p\otimes \rho) \right)
 -H_{\scriptsize{fin}}( \Sigma : {\mathcal A})\\
  \label{balance3}
 &\stackrel{(\ref{condmod})}{=}&H\left(T\,p \right) +H\left(M_2(R\, p\otimes \rho) \right)
 -H_{\scriptsize{fin}}( \Sigma : {\mathcal A})
 \stackrel{(\ref{balance1})}{\ge} H(p)+H(\rho)
 \;.
\end{eqnarray}
Upon the assumption that the conditional action  does not increase the entropy, $H(p)\ge H(T\,p)$,
we further conclude
\begin{equation}\label{balance 4}
 \left|\Delta H(\Sigma) \right| = H(p) -H(T\,p) \stackrel{(\ref{balance3})}{\le}
 H\left(M_2(R\, p\otimes \rho) \right)-H(\rho)
 -H_{\scriptsize{fin}}( \Sigma : {\mathcal A}) = \Delta H({\mathcal A}) -H_{\scriptsize{fin}}( \Sigma : {\mathcal A})
 \;,
\end{equation}
and hence the decrease of entropy of the system $\Sigma$ is (over)compensated by
the increase of entropy of the apparatus ${\mathcal A}$  minus the mutual entropy
between $\Sigma$ and ${\mathcal A}$, similar to the pure case, see (\ref{DeltaH}).

Recall that in our construction of the model the entropy  $H_{\scriptstyle fin} ({\mathcal A})$
can be viewed as the entropy cost of the conditional action including the
measurement and the storage of its result as in the original Szilard principle.
Taking into account a general time evolution of the combined system $\Sigma {\mathcal A}$
given by a doubly-stochastic transition matrix we may formulate this result as a modified S principle:
\setcounter{prin}{0}
\begin{prin} \label{PS}(Modified S principle)
If a conditional action on a system $\Sigma$ is modeled by an interaction with an apparatus ${\mathcal A}$
then the decrease in entropy $\Delta H(\Sigma)$ is at least compensated by the
increase in entropy  $\Delta H({\mathcal A})$ of the apparatus ${\mathcal A}$ minus the mutual information
$H_{\scriptstyle fin} (\Sigma : {\mathcal A})$. $\Delta H({\mathcal A})$ is the entropy increase due to the
conditional action including the measurement on $\Sigma$ and the storage of its result in ${\mathcal A}$.
\end{prin}

The modification of the S principle essentially consists of the inclusion of the mutual information
into the entropy balance.
This is, of course, well known, see e.~g.~\cite{L89}.
In this paper a calculation analogous to ours was used to argue
for the possibility of a measurement without entropy increase as opposed to the S principle.
In our view a model for a reversible measurement (or pure conditional action) in the enlarged system
$\Sigma {\mathcal A}$ implies that the decrease of entropy in $\Sigma$ must be compensated by
an increase of entropy in ${\mathcal A}$ (minus the mutual information) and hence is
rather an argument in favor of the (modified) S principle.

\section{Physical model of the Szilard engine}\label{sec:MSE}

\subsection{General description}\label{sec:GD}

\begin{figure}[ht]
  \centering
    \includegraphics[width=1.0\linewidth]{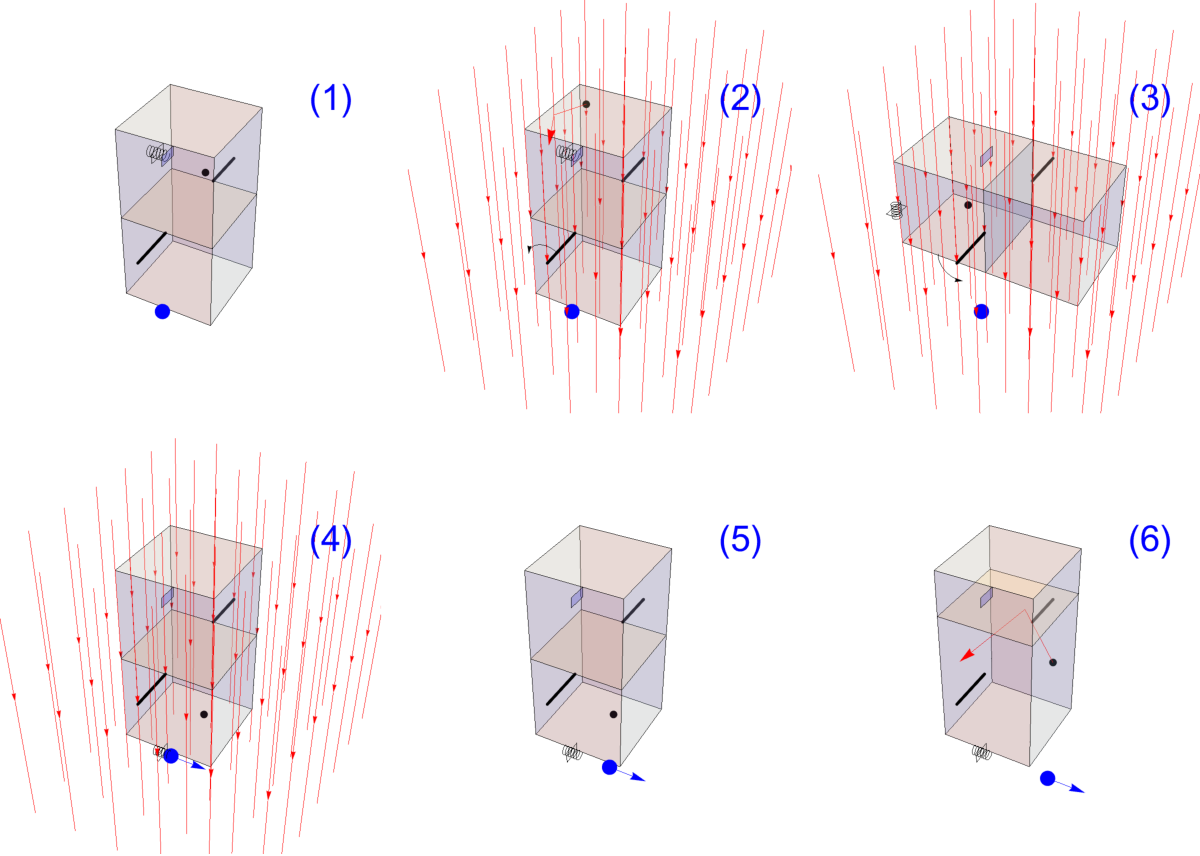}
  \caption{
  Sketch of the ``Szilard pendulum" realizing the conditional action in the case where an
  electrically charged molecule (black dot) is initially
  in the upper box. The pendulum consists of a container with two cube-shaped boxes,
  separated by a piston,
  that can rotate around the $y$-axis (thick black lines, picture (1)).
  Then an electric field is applied (red field lines, picture (2)).
  Small fluctuations cause the pendulum to leave its unstable rest position
  and tilt to the left (picture (3)). Tilting to the right is blocked
  (indicated by a small blue rectangle).
  The pendulum moves downwards until it hits an auxiliary particle (blue dot)
  at its lowest point (Figure (4)), to which it transfers its kinetic energy
  and angular momentum. The elastic collision between the pendulum and
  the auxiliary particle is mediated by a small spring.
  Then the electric field is removed (picture (5)),
  the rotation is blocked and the isothermal expansion of the single-molecule
  gas begins by moving the piston upwards (picture (6)).
  }
  \label{FIGSBA}
\end{figure}

We shortly recapitulate the general description of the Szilard engine given in Section \ref{sec:P}.
The container is placed in a vertical gravitational field and, after the
insertion of the partition and a position measurement of the molecule,
rotated about $180^\circ$ if the molecule was found in the upper box.
Then the partition is unlocked and serves as a piston that is
moved upwards by collisions with the molecule.
The condition that enables a slow isothermal expansion is
\begin{equation}\label{condexp}
  \kB T \sim \mu\, g\, a
  \;,
\end{equation}
where $\kB T$ denotes a typical thermal energy of the molecule, $\mu$ is the mass of the piston,
usually large compared to the mass $m$ of the molecule and $a$ is the height of the lower box
with volume $V_1=A\,a$.
This follows from the force balance $p A \sim \mu g$ and the ideal gas law $p\, A\, a= N \,\kB \, T$
($p$: pressure)
which is assumed to hold also for the one-particle gas with $N=1$. The condition
\begin{equation}\label{condmass}
 \mu \gg m
\end{equation}
guarantees that the probability density of the molecule w.~r.~t.~position is practically constant
and effects due to the barometric formula can be neglected.

In the next step,
we consider a mechanical implementation of the conditional action
just described, see Figure \ref{FIGSBA}.
For the sake of simplicity, let us assume that the two partial
volumes have the shape of cubes with edge length $a$ and
that the inertial tensor $\theta$ of the container will be approximately
diagonal w.~r.~t.~the chosen coordinate system
having the eigenvalues (inertial moments) $\theta_x,\theta_y,\theta_z$.
We further assume that the molecule is ionized and hence equipped with a charge, say $q>0$, and
that a constant electric field with a component $E>0$ in the direction of the gravitational field
is applied after the rotation of the container is unlocked. This can be
achieved by slowly moving a large plate capacitor (not shown in Figure \ref{FIGSBA}) into the region of the container.
If the molecule is located in the upper box of the container
that can rotate around the $y$-axis, it represents a pendulum in its unstable rest position.
The reflections of the molecule at the walls of the container cause fluctuations
into the $y$-direction which shift the pendulum beyond its rest position and start an
oscillation with an angular amplitude of approximately $180^\circ$.
To simplify the calculations we will assume that the pendulum can only tilt to the left
since the other direction is blocked.
Without detailed calculations we will conclude that the motion of the pendulum is only
slightly superimposed by thermal fluctuations as long as the condition
\begin{equation}\label{condpend}
  \kB T \ll (m\, g+q\, E)\, a \stackrel{(\ref{condexp})(\ref{condmass})}{\sim} q\, E\, a
\end{equation}
holds.

This condition also implies that in thermal equilibrium the spatial probability density
will be concentrated at the bottom of the box in which the molecule is located.
This is in sharp contrast to the case of pure gravity,
where the spatial probability density within the box is practically constant.

When the pendulum has reached its lowest point, it has an approximate kinetic energy of
$E_{\scriptstyle{kin}}=q\, E\, a = \frac{1}{2} \theta_y \omega^2$ and a $y$-component of the angular momentum
$L_y=\theta_y \omega$, where $\omega$ denotes its angular velocity.

We place another auxiliary particle with mass $M$
in the plane $z=-a$ so that the pendulum collides with it elastically at its lowest point
and after the collision the auxiliary particle moves frictionless along the
line $z=-a, y=-a/2$ with the speed $v$.
In order that the pendulum completely transfers its kinetic energy and angular momentum to the auxiliary particle
the following equations have to be fulfilled:
\begin{equation}\label{completely}
q\, E\, a = \frac{1}{2} \theta_y \omega^2= \frac{M}{2} v^2, \quad \theta_y \omega = M\,v\,a
\;.
\end{equation}
The positive solutions to these equations read
\begin{equation}\label{solu}
M=\frac{\theta_y}{a^2},\quad  v =  a^{3/2} \sqrt{\frac{2 q E}{\theta_y }}, \quad \mbox{and }
\omega =  \sqrt{\frac{2 q a E}{\theta_y }}
 \;.
\end{equation}
We will choose the mass of the auxiliary particle correspondingly.

If the molecule initially was in the lower box, the auxiliary particle would remain at rest.
Thus the state of the auxiliary particle after the conditional action can be viewed as
a record of the position measurement of the molecule.
Note that in contrast to the general description in Section \ref{sec:P} the memory has only two
states since the state $0$ (the auxiliary particle being at rest) is also used as the default value.
But this does not change the entropy balance.

After the Szilard pendulum has reached its lower rest position the
rotations around the $y$-axis are blocked again and the electric field
is removed so as not to prevent the subsequent expansion.

It is obvious that our realization of the conditional action is not a pure one.
If the molecule starts in a pure state corresponding to a small cell of its phase space belonging, say, to the upper box,
then at the end of the conditional action it will be in a mixed state
corresponding to a uniform distribution in the lower box and a thermal distribution in momentum space.
This is the reason why we have extended our general entropy balance to the non-pure case
in subsection \ref{sec:MGA}.

For the energy balance, we note that the electrostatic potential energy
$q\, E\, a$ of the molecule is completely converted into kinetic energy of the
auxiliary particle. This energy must be paid for by the external energy store
because this is the amount by which the work required to move the plate capacitor
into the domain of the container exceeds the work gained by removing the capacitor.
Naively, one might think that both kinds of work disappear, since the direction of movement
of the capacitor is perpendicular to the electric field lines.
However, this neglects the deformation of the constant field outside the capacitor
and the additional deformation caused by charge displacements on the plates of the capacitor induced
by the charged gas molecule.

As in the original work by Szilard \cite{S29}, the expansion of the single-particle
gas is carried out isothermally. For example, we can assume the walls of the container
to be a heat bath with a practically constant temperature $T$.

In our approach, we have to consider the heat bath as a finite system,
while the usual idealization of heat baths makes the transition to infinite systems.
This is not a real problem, as we will also make analogous approximation assumptions
concerning the state of the heat bath before and after the Szilard process.

In addition, we can also consider a physical implementation of the process
that resets the memory to its default value.
However, this is more of a playful consideration that has no bearing on the rest of this work.
To this end, we consider a second one-molecule-gas consisting only of the auxiliary particle enclosed
in another large container $C_2$. By adiabatic expansion of $C_2$,
we cool this one-particle-gas to an average velocity corresponding to a low temperature $T_0 \ll T$.
In the second step, this gas is isothermally (at temperature $T_0$) compressed
to a small volume that corresponds approximately to the phase space cell for the
auxiliary particle at the beginning of the Szilard process.
Based on this construction, the Szilard process could be interpreted not as a PM2,
but as a Carnot-like cyclic process operating between two heat baths of temperature $T$ and $T_0$.

This completes the description of the slightly modified Szilard engine and the
physical realization of the conditional action.

\subsection{Entropy balance}\label{sec:EB}

A detailed description of the Szilard engine during the conditional action phase
by a mechanical many-body system including the plate capacitor and various
braking devices to block or release certain movements of the container is of course not possible.
Instead, we will make plausible assumptions about the probabilities of the
various states at the beginning and end of this phase.
The calculations are simplified due to the fact that there is no Gibbs paradox for $N=1$.

First, the phase space of the entire many-body system is divided into $N$ sufficiently small cells,
which are occupied with the probabilities $p_i,\; i=1,\ldots, N,$.
We concentrate on the cells for the molecule and the auxiliary particle,
as we can assume that the remaining probabilities do not change
between the beginning and end of the conditional action phase.
We will use the following notation. The spatial domain of the lower box is divided into $N_s$ cells
of equal size with indices $\ell\in {\mathcal L}$, analogously for the upper box with $N_s$ cells
and indices $u\in {\mathcal U}$. The momentum space of the molecule is divided into cells
of equal size (hence introducing a momentum cutt-off) with indices $p\in {\mathcal P}$ and mean momentum ${\mathbf p}$.
Only two cells of the phase space for the auxiliary particle
will be singled out, since the other cells are not occupied: The cell with index $v_0$ of the
auxiliary particle at rest and the cell with index $v_1$ corresponding to the moving
particle after the interaction with the Szilard pendulum, see Figure \ref{FIGSBA},
at a fixed time corresponding to the end of the conditional action phase.
The following assumptions about the corresponding probabilities are always understood as
approximate identities although they will be denoted by equality signs.
At the beginning the probabilities of the molecule's state will be
\begin{equation}\label{pinimol}
 p_{\scriptstyle{ini}}(\ell,p)= p_{\scriptstyle{ini}}(u,p) = \frac{1}{2 N_s} \pi(p)
 \;,
\end{equation}
for all $\ell\in {\mathcal L},\,u\in {\mathcal U}$ and $p\in {\mathcal P}$, where $\pi(p)$
is the thermal probability
\begin{equation}\label{pipdef}
 \pi(p):= \frac{1}{Z} \exp\left( -\frac{{\mathbf p}^2}{2 m \kB T}\right)
 \;,
\end{equation}
and $Z$ is a normalization factor. As noted above, we have neglected the small $z$-dependence
of the probability due to the barometric formula. The probabilities of the auxiliary particle
are
\begin{equation}\label{pauxini}
 p_{\scriptstyle{ini}}(v_0)=1,\quad \mbox{and}\quad  p_{\scriptstyle{ini}}(v_1)=0
 \;.
\end{equation}

Next we consider the probabilities at the end of the conditional action.
As a result of the conditional $180^\circ$ rotation, the two boxes have the same positions as before,
but are swapped with a probability of $1/2$.
According to the isothermal character of the process the thermal probability of the molecule
will be practically unchanged. Hence
\begin{equation}\label{pfinmol}
 p_{\scriptstyle{fin}}(\ell,p)= \frac{1}{ N_s} \pi(p),\quad \mbox{and}\quad p_{\scriptstyle{fin}}(u,p)= 0
 \;,
\end{equation}
for all $\ell\in {\mathcal L},\,u\in {\mathcal U}$ and $p\in {\mathcal P}$.
For the auxiliary particle it holds that
\begin{equation}\label{pauxfin}
 p_{\scriptstyle{fin}}(v_0)= p_{\scriptstyle{fin}}(v_1)=\frac{1}{2}
 \;.
\end{equation}
Note that the probability distributions of the molecule and the auxiliary particle are independent,
although the state of the auxiliary particle encodes the information of the initial position
in the upper or lower box of the molecule. But this information has been lost in the coarse-grained
description of the molecule and the remaining part of the Szilard engine.

We next calculate the entropies. 

The symbol $\Sigma$ refers to the one-molecule gas,
and ${\mathcal A}$ to the remaining parts of the Szilard engine
including the heat bath, as explained in Section \ref{sec:P}.
Part of ${\mathcal A}$ consists of the auxiliary particle denoted by
${\mathcal A}_{\scriptstyle{aux}}$, which acts as a memory,
while all the rest of ${\mathcal A}$ is captured by the symbol ${\mathcal A}_{\scriptstyle{r}}$.

We obtain
\begin{equation}\label{HiniSigma}
H_{\scriptstyle{ini}}(\Sigma)=-\sum_{\ell\in{\mathcal L}}\frac{1}{2 N_s} \log \frac{1}{2 N_s} -
\sum_{u\in{\mathcal U}}\frac{1}{2 N_s} \log \frac{1}{2 N_s} + H_p = \log 2 +\log N_s + H_p
\;,
\end{equation}
where $H_p$ is the part of the entropy corresponding to the thermal probability distribution, namely
\begin{equation}\label{Hp}
 H_p= - \sum_{p\in {\mathcal P}} \left( -\frac{{\mathbf p}^2}{2 m \kB T}- \log Z \right)
 \, \frac{1}{Z}\,\exp\left( -\frac{{\mathbf p}^2}{2 m \kB T}\right)
  \;.
\end{equation}
The entropy of the auxiliary particle vanishes and thus we obtain for the total entropy
\begin{equation}\label{Htotini}
  H_{\scriptstyle{ini}}(\Sigma ,\,{\mathcal A})= H_{\scriptstyle{ini}}({\mathcal A}_{\scriptstyle{r}})+ H_p + \log N_s +\log 2
  \;,
\end{equation}
assuming that the mutual informations  $H_{\scriptstyle{ini}}(\Sigma : {\mathcal A}_{\scriptstyle{r}})$
and $H_{\scriptstyle{ini}}\left({\mathcal A}_{\scriptstyle{aux}} : {\mathcal A}_{\scriptstyle{r}}\right)$ can be neglected.

Analogously, we obtain for the entropies after the conditional action
\begin{equation}\label{HfinSigma}
H_{\scriptstyle{fin}}(\Sigma)=-\sum_{\ell\in{\mathcal L}}\frac{1}{N_s} \log \frac{1}{N_s}+ H_p = \log N_s + H_p
\;.
\end{equation}
This time the entropy of the auxiliary particle becomes
\begin{equation}\label{Hauxfin}
  H_{\scriptstyle{fin}}({\mathcal A}_{\scriptstyle{aux}})= -\frac{1}{2}\log \frac{1}{2}
 -\frac{1}{2}\log \frac{1}{2}= \log 2
\;,
\end{equation}
and hence
\begin{equation}\label{Htotfin}
  H_{\scriptstyle{fin}}(\Sigma, \,{\mathcal A})= H_{\scriptstyle{fin}}({\mathcal A}_{\scriptstyle{r}})+ H_p + \log N_s +\log 2
  \;,
\end{equation}
The electrostatic and gravitational potential energy difference between the molecule in the upper box and in the lower one has been
completely transferred into kinetic energy of the auxiliary particle. Hence the energy of
${\mathcal A}_{\scriptstyle{r}}$ is unchanged and it is sensible to assume that also
\begin{equation}\label{HAr}
  H_{\scriptstyle{ini}}({\mathcal A}_{\scriptstyle{r}}) = H_{\scriptstyle{fin}}({\mathcal A}_{\scriptstyle{r}})
  \;.
\end{equation}
This means that
$H_{\scriptstyle{ini}}(\Sigma, \,{\mathcal A})= H_{\scriptstyle{fin}}(\Sigma, \,{\mathcal A})$ and, due to the vanishing mutual information,
the entropy decrease $\Delta H(\Sigma)= -\log 2$ of the gas is exactly compensated
by the entropy increase $\Delta H({\mathcal A}_{\scriptstyle{aux}})= \log 2$ of the
auxiliary particle, in accordance with the modified S principle.

The following isothermal expansion leads to an entropy flow from the heat bath to the gas and
does not change the total entropy.

\subsection{Entropy balance of a modified Szilard engine}\label{sec:EBM}

\begin{figure}[ht]
  \centering
    \includegraphics[width=1.0\linewidth]{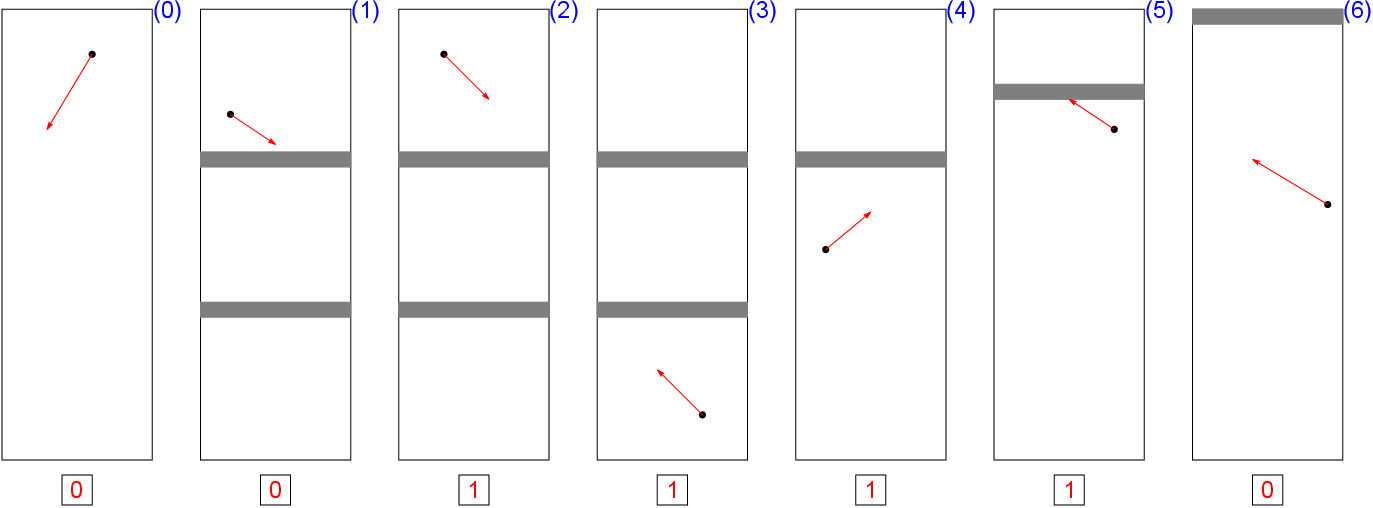}
  \caption{
  Sketch of the different phases of a modified three-boxed Szilard engine cycle.
  The whole container is put into a vertical gravitational field. The walls of the container serve as a heat bath
  for the one-molecule gas.  The memory is set to its default value $0$ (picture (0)).
  Two partitions are inserted which divide the container into three boxes of equal volume (picture (1)).
  Then a position measurement is carried out and the result is stored in the memory (picture (2)).
  We will only show the case where the molecule is initially in the upper box and
  the corresponding measurement result is $1$. Only in this case the container is rotated about $180^\circ$
  such that now the molecule is in the lower box (picture (3)).
  The lower partition is removed (picture (4)). The upper partition is unlocked and serves as a piston that is moved upwards
  by the collisions with the molecule (picture (5)) until it reaches its final position.
  Then the   memory is erased and set to the default value $0$ (picture (6)).
  }
  \label{FIGP3}
\end{figure}

Our modification of the Szilard principle consisted of the additional consideration of the mutual information.
It is therefore somewhat unfortunate that the analysis of the Szilard engine
makes this modification appear superfluous, because the mutual information disappears here.
Therefore, we now consider another modification of the Szilard engine in which the
mutual information does {\it not} disappear after completion of the conditional action,
see Figure \ref{FIGP3}.

At the beginning of the process, {\it two} partitions are inserted this time,
which divide the container into three equal partial volumes, say,
into three cubes with edge length $a$. 

As in the Szilard machine with two boxes, an electric field is
applied which exerts a downward force on the charged molecule
and can cause the Szilard pendulum to rotate macroscopically.
But this only happens if the molecule is in the upper box. Only then
will the  pendulum tilt to the left and, as in the previous case, hit the auxiliary particle.
If the molecule is in the middle box the pendulum will remain in a stable rest
position because, due to the strong electric field,
the molecule will be found near the bottom of the box with overwhelming probability,
see the analogous remark in subsection \ref{sec:GD}.

The state of the auxiliary particle after the conditional action
thus represents the result of the measurement of whether the molecule
was originally either in the upper box or in one of the other two boxes;
this is a measurement with {\it two} possible outcomes, not three.
After the pendulum has come to a standstill, we obtain a state in which the
molecule is in the lower box with a probability of $2/3$
and in the middle box with a probability of $1/3$.
In contrast to the two-boxed engine, this state is still correlated
with the state of the auxiliary particle with $p(v_0)=2/3$ and $p(v_1)=1/3$.

The detailed calculation, analogous to the previous case, shows that
\begin{equation}\label{Hinisigmamod}
  H_{\scriptstyle{ini}}(\Sigma) = \log 3 +\log N_s + H_p
  \;,
\end{equation}
where $N_s$ is again the number of cells belonging to each of the three boxes,
denoted by  ${\mathcal L}$, ${\mathcal M}$ and ${\mathcal U}$.
Moreover,
\begin{eqnarray}
\label{Hfinsigmamoda}
   H_{\scriptstyle{fin}}(\Sigma)  &=& -\sum_{\ell\in{\mathcal L}} \frac{2}{3 N_s} \log \frac{2}{3 N_s}
    -\sum_{m\in{\mathcal M}} \frac{1}{3 N_s} \log \frac{1}{3 N_s} +H_p\\
  \label{Hfinsigmamodb}
   &=& \frac{2}{3}\left(\log 3 -\log 2 +\log N_s \right)+ \frac{1}{3}\left(\log 3 +\log N_s \right)+H_p\\
   \label{Hfinsigmamodc}
   &=& \log 3 -\frac{2}{3} \log 2 +\log N_s +H_p
   \;.
\end{eqnarray}
This results in an entropy decrease of the gas given by its absolute value
\begin{equation}\label{deltaHsigmamod}
  \left| \Delta H(\Sigma)\right| = \frac{2}{3} \log 2 = 0.462098\ldots
  \;.
\end{equation}
The entropy increase of the auxiliary particle is obtained as
\begin{eqnarray}
\label{deltaHauxmoda}
  \Delta H({\mathcal A}_{{\scriptstyle{aux}}})&=& H_{\scriptstyle{fin}}({\mathcal A}_{{\scriptstyle{aux}}})
  = -\frac{2}{3} \log\frac{2}{3} -\frac{1}{3} \log\frac{1}{3}
 \\
 \label{deltaHauxmodb}
   &=& \frac{2}{3}\left( \log 3 -\log 2\right)+ \frac{1}{3} \log 3 =\log 3 -\frac{2}{3}\log 2=0.636514\ldots
   \;.
\end{eqnarray}

It remains to calculate the mutual information $H_{\scriptstyle{fin}}(\Sigma: {\mathcal A}_{\scriptstyle{aux}})$.
We write down the joint probabilities
\begin{eqnarray}
\label{pSigmaAauxa}
 p(\ell,p,v_0) &=& \frac{1}{3 N_s}\,\pi(p) \\
 \label{pSigmaAauxb}
 p(\ell,p,v_1) &=& \frac{1}{3 N_s}\,\pi(p) \\
  \label{pSigmaAauxc}
 p(m,p,v_0) &=& \frac{1}{3 N_s}\,\pi(p) \\
 \label{pSigmaAauxd}
 p(m,p,v_1) &=& 0
 \;,
\end{eqnarray}
for all $\ell\in{\mathcal L}$, $\ell\in{\mathcal L}$ and $p\in{\mathcal P}$.
This results in the total entropy of
\begin{eqnarray}
\label{HtotSigmaAa}
  H_{\scriptstyle{fin}}(\Sigma,\, {\mathcal A}_{\scriptstyle{aux}}) &=&
  -\sum_{\ell\in{\mathcal L}} \frac{1}{3N_s}\log\frac{1}{3 N_s}
  -\sum_{\ell\in{\mathcal L}} \frac{1}{3N_s}\log\frac{1}{3 N_s}
  -\sum_{m\in{\mathcal M}} \frac{1}{3N_s}\log\frac{1}{3 N_s} +H_p\\
  \label{HtotSigmaAb}
    &=& 3\times\frac{1}{3}\left( \log 3 +\log N_s\right)+   H_p\\
    \label{HtotSigmaAc}
    &=& \log 3 + \log N_s +H_p
    \;.
\end{eqnarray}
Hence the mutual information assumes the value
\begin{eqnarray}
\label{correnta}
  H_{\scriptstyle{fin}}(\Sigma: {\mathcal A}_{\scriptstyle{aux}}) &=&
  H_{\scriptstyle{fin}}(\Sigma) + H_{\scriptstyle{fin}}({\mathcal A}_{\scriptstyle{aux}})-
  H_{\scriptstyle{fin}}(\Sigma,\, {\mathcal A}_{\scriptstyle{aux}}) \\
  \label{correntb}
   &\stackrel{(\ref{Hfinsigmamodc})(\ref{deltaHauxmodb})}{=}&
   \left(\log 3 -\frac{2}{3} \log 2 +\log N_s +H_p \right) +
   \left(\log 3 -\frac{2}{3}\log 2 \right) -
   \left( \log 3 + \log N_s  +H_p \right)\\
    \label{correntc}
    &=& \log 3-\frac{4}{3} \log 2  = 0.174416\ldots >0
    \;.
\end{eqnarray}

We conclude that the entropy decrease of the gas $\Delta H(\Sigma)$ is
exactly compensated by the entropy increase of the auxiliary particle
$\Delta H({\mathcal A}_{{\scriptstyle{aux}}})$ minus the  mutual information
$ H_{\scriptstyle{fin}}(\Sigma: {\mathcal A}_{\scriptstyle{aux}})$:
\begin{eqnarray}\label{overa}
 \left( \Delta H({\mathcal A}_{{\scriptstyle{aux}}})- H_{\scriptstyle{fin}}(\Sigma: {\mathcal A}_{\scriptstyle{aux}})\right)-
 \left|\Delta H(\Sigma)\right|
 \stackrel{(\ref{deltaHsigmamod})(\ref{deltaHauxmodb})(\ref{correntc})}{=}&&\\
 \label{overb}
 \left(\left(\log 3 -\frac{2}{3}\log 2  \right)- \left(\log 3 -\frac{4}{3}\log 2 \right)\right) -\frac{2}{3}\log 2 =0
\;.
\end{eqnarray}
This result is in accordance with the modified S principle if we assume, as for the two-boxed Szilard engine,
that the remaining part ${\mathcal A}_{{\scriptstyle{r}}}$ of the engine does not contribute to
the entropy balance (except for the isothermal expansion phase).

Although the focus of our investigation is on the entropy balance during
the conditional action, we will briefly discuss the further process for the three-boxed Szilard engine.
After the conditional action described above, the lower partition is removed,
increasing the total entropy. Finally, the upper partition is unlocked,
turning it into a piston that is pushed upwards by the molecule until it reaches its highest position,
see Figure \ref{FIGP3}.

After the amount $\mu \, g \, a$ of the potential energy of the piston has been transferred to an external energy store,
the two partitions are re-inserted and the entire process could possibly be continued as a cyclic process.
However, this would also require the auxiliary particle to be returned to its original state.
This makes it clear that also our realization of the three-boxed Szilard engine does not lead to a PM2.
Nevertheless, we want to discuss the question of whether it is even conceivable
as a proposal for a PM2.
The difference to the two-boxed Szilard engine is the additional entropy production
by pulling out the lower partition wall before the isothermal expansion.
This entropy production must be compared with the entropy loss of an alleged PM2.

In fact, a short calculation shows that by pulling out the lower partition wall,
the entropy of the gas + auxiliary particle system  
$\Sigma\,{\mathcal A}_{\scriptstyle{aux}}$
increases by the amount $\Delta \widetilde{H}=\frac{1}{3} \log 2$,
taking into account the vanishing mutual information at the end of the process.
On the other hand,  due to the isothermal expansion the entropy of the heat bath
decreases by the amount $\Delta H'=\log 3/2$, which would have to be interpreted
as the entropy loss of a cycle of the alleged PM2.
Obviously,
\begin{equation}\label{compare}
\Delta H' = \log 3 - \log 2 =  0.405465\ldots   >  \Delta \widetilde{H}=\frac{1}{3} \log 2 = 0.231049\ldots
\;,
\end{equation}
and thus the three-boxed Szilard engine would still be a candidate of a PM2
if one neglects the entropy increase of the auxiliary particle that serves as information storage for the measurement result.

\section{Discussion and Summary}\label{sec:SUM}

In contrast to quantum theory, the theoretical concept of ``measurement"
plays a less central role in classical theories. This may be due to the fact that,
at least approximately, classical theories allow for the idea that measurement
merely reveals pre-existing facts. In contrast, quantum theory suggests that
these facts are partially created by the act of measurement itself.
A notable exception to this observation is the theoretical role of measurement
in the entropy balance of thought experiments such as Maxwell's demon or Szilard's engine.
In this context, the increase in entropy due to measurement and the storage of measurement
results has been used to counteract the decrease in entropy caused by external intervention (S principle).
Criticism of the S principle has highlighted the possibility of reversible measurements
and instead emphasized entropy production during the erasure of measurement results
as a means to uphold the second law of thermodynamics (LB principle).
Without such erasure, the Szilard process, for example, could not function as a cyclic process.

It seems that the majority of the physical community is more inclined towards
the LB principle than the S principle.  In this article, we argue that this is unfounded.
In order to substantiate this thesis, we have carried out a detailed analysis
of the individual phases of the Szilard process.
Various slightly different versions of this process exist,
but we have selected a particular realization that aligns well with our approach.
Our method is based on the observation that the entropy reduction in the
system under consideration, such as the one-particle gas in Szilard's engine,
is not directly caused by the measurement but rather by the external intervention
dependent on the measurement outcome. We refer to such an intervention as
``conditional action," following earlier work \cite{S20}, \cite{S21}.

In Section \ref{sec:GTCA}, we develop a simplified theory of classical pure conditional actions.
The core insight of this theory is that a classical conditional action merges states,
reducing entropy in the system $\Sigma$ due to the decreased number of occupied states.
A physical realization of a conditional action involves the time evolution
of a larger system $\Sigma {\mathcal A}$, where ${\mathcal A}$ represents
the apparatus responsible for measurement and external intervention.
Since the time evolution of $\Sigma {\mathcal A}$ does not reduce total entropy,
this setup inherently provides the necessary compensation for entropy reduction in $\Sigma$.
This raises concerns about circular reasoning, similar to objections raised against the LB principle.
However, the present discussion is not about the validity of the second law of thermodynamics
but rather about a consistent analysis of a thought experiment -
just as one would try to find hidden issues in a proposal for a {\it perpetuum mobile}
of the first kind using classical mechanics, implicitly assuming energy conservation.

Thus, the entropy balance in the general case can be described as follows:
\begin{itemize}
  \item {\bf Measurement Phase}: A one-to-one relationship is established
  between certain states of $\Sigma$  and states of  ${\mathcal A}$.
  This increases the entropy of ${\mathcal A}$, which is compensated
  by an equal increase in mutual information between $\Sigma$ and ${\mathcal A}$.
  Since the measurement can be carried out reversibly,
  criticism of the S principle is formally correct but
  ultimately irrelevant, as no entropy reduction occurs in  $\Sigma$.
  \item {\bf Conditional Action Phase}:
  The reduction in entropy occurs only in this phase and is simultaneously
  compensated by the (partial) elimination of mutual information between $\Sigma$ and ${\mathcal A}$.
  As the states merge, their correlation with memory states is (partially) lost.
\end{itemize}

This finding challenges the LB principle, which claims that compensation
occurs generally only through memory content erasure.
However, we emphasize that in special cases, a different picture may emerge:
 memory content erasure could coincide with conditional action,
 thus restoring the LB principle. In such scenarios,
 entropy reduction would only be observed at the end of the process
 and would occur not in $\Sigma$ but in ${\mathcal A}$.
In the general case, we conclude that instead of the LB principle,
a modified S principle applies (see Section \ref{sec:MGA}).
This principle explains that the entropy decrease in $\Sigma$
during the combined phase ``measurement + conditional action" is compatible
with the 2nd law by offsetting it against the entropy increase in the apparatus ${\mathcal A}$,
after subtracting the remaining mutual information.
The phases of ``measurement" and ``conditional action" are considered together,
since some physical realizations, such as that of the Szilard engine in this paper,
do not allow them to be clearly separated.

Beyond illustrating the general theory,
a concrete realization of Szilard's engine helps bridge the gap between the abstract
theory of conditional actions and the physical thought experiment.
We propose positioning the Szilard engine's container in a gravitational field
so that the work released during isothermal expansion directly
increases the potential energy of the (now non-massless) piston.
This setup only works if the gas particle starts in the lower box.
If it is in the upper box, the container must be flipped,
which constitutes a conditional action and presupposes a coarse position measurement.

As a physical implementation of the conditional action, we suggest electrically charging the gas particle
and using the adiabatic transport of a plate capacitor to place it in an electric field.
If the container is rotatable, it functions as a pendulum that enacts the conditional rotation.
After a $180^\circ$ rotation, the pendulum must decelerate,
which we achieve through an elastic collision with an auxiliary particle.
The auxiliary particle's momentum serves as a record of the gas particle's initial position (upper or lower box).
The electric field is then removed for the subsequent isothermal expansion.
A detailed entropy balance of this Szilard engine realization supports the
hypotheses formulated in Section \ref{sec:P}.
However, in this case, the mutual information between $\Sigma$ and ${\mathcal A}$ vanishes.
To demonstrate that this is not always the case and
that our modification of the S principle remains relevant,
we briefly discuss a three-boxed variant of the Szilard engine in Section \ref{sec:EBM}.
In this variant, the ``Szilard pendulum" tips only if the gas particle
starts in the uppermost of the three boxes.
After the conditional action, the gas particle is in the lower box
with a probability of $2/3$ and in the middle box with a probability of $1/3$.
However, if the auxiliary particle has positive momentum, the gas particle must be in the lower box.
Therefore, in this case, mutual information does not disappear,
confirming that our modification of the S principle is necessary.

We hope that this work will help to clarify the debate triggered by Maxwell's demon and the Szilard engine.

\section*{Acknowledgment}
We would like to thank Hans-Werner Sch\"urmann for his interest in our work and valuable references to the literature.

\end{document}